\documentclass[aps,prb,twocolumn,preprintnumbers,amsmath,amssymb]{revtex4}
\usepackage{graphicx}
\begin{document}
\title{Inexhaustible physics of the helical magnet MnSi: field evolution of the magnetic phase transition inferred from ultrasound studies}

\author{A.E. Petrova}
\author{S.M. Stishov}
\email{sergei@hppi.troitsk.ru}
\affiliation{Institute for High Pressure Physics of RAS, Troitsk, Russia}

\begin{abstract}
The longitudinal and transverse ultrasound speeds and attenuation were measured in a MnSi single crystal in the temperature range of 2 -- 40 K and magnetic fields to 7 Tesla. The magnetic phase diagram of MnSi in applied magnetic field appears to depend on the experimental setups, which is related to a difference in demagnetization factors arising due to the disc shape of the sample.

The magnetic phase transition in MnSi in zero magnetic field is signified by a quasi discontinuity in the $c_{11}$ elastic constant, which varies significantly with magnetic field. It is notable that the region where the $c_{11}$ discontinuity almost vanishes closely corresponds to the extent of skyrmion phase along the magnetic to paramagnetic transition. This implies that the $c_{11}$ elastic constant is almost continuous through the transition from the skyrmion to paramagnetic phases. A recovery of the discontinuity of $c_{11}$ and enhanced sound absorption occur at the crossing of the phase transition line and the line of minima in $c_{11}$. The powerful fluctuations at the minima of $c_{11}$ make the mentioned crossing point similar to a critical end point, where a second order phase transition meets a first order one.

The skyrmion domain in the case of a "perpendicular" setup with a smaller demagnetization factor has a reduced temperature range, which suggests that the magnetic field inhomogeneity plays an important role in the skyrmion occurrence and, hence, opens a way of skyrmion manipulation.

The small anisotropy of the shear moduli in the (001) plane found in the “parallel” setup is most probably also caused by the magnetic field inhomogeneity, which distorts the hexagonal symmetry of the skyrmion crystal.

\end{abstract}
\maketitle

\section{Introduction}
\begin{figure}[htb]
\includegraphics[width=80mm]{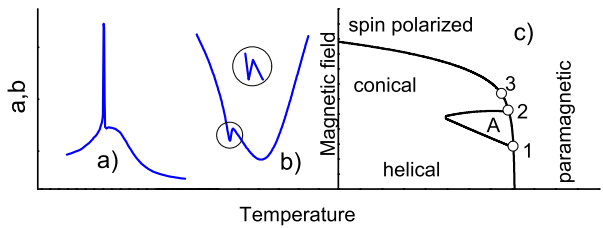}
\caption{\label{fig1} (Color online) Schematic view of the behavior of a) the heat capacity, thermal expansion coefficient, temperature coefficient of electrical resistivity, ultrasound attenuation and b) elastic moduli at the phase transition in MnSi c) schematic magnetic phase diagram of MnSi.}
\end{figure}

As is well known, the intermetallic compound MnSi experiences a phase transition to a spin ordered phase at $\sim$29 K~\cite{1,2}. The magnetic order in MnSi at zero magnetic field is identified as a long period ferromagnetic spiral or a helical spin structure with a helical twist caused by the Dzyaloshinski--Moria interaction~\cite{3}. The helical-paramagnetic phase transition at zero magnetic field, thought to be of second order, is now recognized as a first order transition. Some experimental data in favor of a first order transition in MnSi were obtained long ago and were subsequently supported by theoretical analysis~\cite{4}. But these indications were mostly ignored, possibly because of potential contradiction to the attractive concept of the existence of a tricritical point in MnSi at high pressure~\cite{5}. However, new data on the heat capacity, thermal expansion, electrical resistivity, the elastic properties and from neutron scattering clearly evidenced the first order nature of the phase transition in MnSi~\cite{6,7,8,9}. Fig.~\ref{fig1} illustrates the behavior of various physical quantities at the phase transition in MnSi. Sharp peaks on the low temperature side of shallow maxima of fluctuation origin (Fig.~\ref{fig1}a) are characteristic for the heat capacity, thermal expansion coefficient, temperature coefficient of the electrical resistivity, and the ultrasound attenuation~\cite{6,7,8,9}. Elastic moduli near the phase transition in MnSi demonstrate small quasi discontinuities on the low temperature side of the fluctuation minima (Fig.~\ref{fig1}b)~\cite{8}. This discontinuity clearly evidences the first order character of the magnetic phase transition in MnSi. Of course, a discontinuous change of elastic constants may be observed also at some second order phase transitions but that kind of transition can not be accompanied by a diverging heat capacity as it does in case of MnSi~\cite{6,7}.

Currently the Brazovski theory of fluctuation-induced first order phase transitions has become very popular in the analysis of the MnSi system~\cite{10,11}.

On application of magnetic fields the helical spin structure of MnSi transforms to a conical one and finally to a spin polarized paramagnetic structure~\cite{12}.A rough scheme of the $H-T$ phase diagram of MnSi is shown in Fig.~\ref{fig1}c. A small pocket of a so-called A-phase exists in the vicinity of the transition to the paramagnetic phase at magnetic fields $\sim$0.1-0.2 T~\cite{13}. Recent neutron scattering studies have revealed features of the skyrmion spin structure of the A-phase~\cite{14}. Heat capacity studies of the magnetic phase transition in magnetic fields permit the authors of Ref.~\cite{15} to claim the existence of a tricritical point at $\sim$0.4 T and $\sim$28 K on the $T-H$ equilibrium line (point 3 in Fig.~\ref{fig1}c), corresponding to the coexistence of the magnetic and paramagnetic phase of MnSi.  The latest ultrasound studies have revealed a rather sharp change of the elastic behavior of MnSi at the phase line separating the skyrmion domain from the conical phase~\cite{16}. However, it is still not quite clear what the nature of the phase transitions from the skyrmion crystal to the paramagnetic and conical phases is. Or, equally, what are the characteristics of the crossing points 1,2 (Fig.~\ref{fig1}c) , where the helical/conical phase line enters and exits the skyrmion domain.

The point is that depending on the nature of the phase boundaries these crossing points could be simple triple points, or multicritical points, or critical end points. The whole situation crucially depends on the nature of the phase transition from the skyrmion phase or phase A to the paramagnetic phase. Indeed it would not be surprising if the nature of the indicated phase line differs somehow from those of both the helical and conical transitions to the paramagnetic phase.

To resolve these issues we performed ultrasound studies of MnSi in magnetic fields 0 -- 7 T and temperature 2 -- 40 K with an emphasis on the magnetic and A-phase transitions.

As discussed above, elastic moduli at the phase transition in MnSi demonstrate small quasi discontinuities on the low temperature side of the fluctuation minima (Fig.~\ref{fig1}b), which can be used for identifying the character of the phase transition~\cite{8}.

\begin{figure}[htb]
\includegraphics[width=80mm]{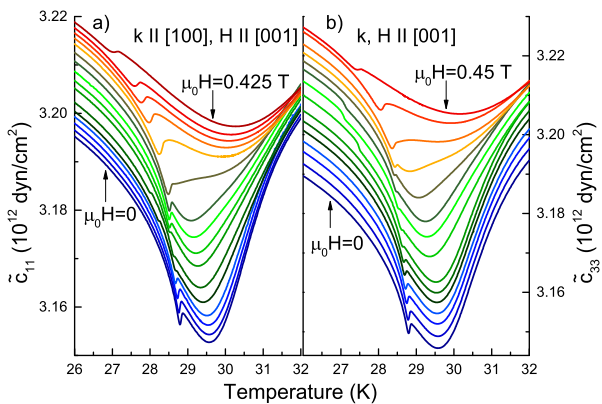}
\caption{\label{fig2} (Color online) Temperature dependence of the elastic moduli $\widetilde{c}_{11}$ and $\widetilde{c}_{33}$ at the phase transition in MnSi (the data are shown with offsets for better viewing).} To avoid confusion we use the tilde sign to mark the elastic moduli to distinguish them from the constants of real tetragonal symmetry.
 \end{figure}

\section{Experimental}
In the course of the ultrasound studies of MnSi, quite a number of runs were performed using digital pulse-echo techniques~\cite{17}. A disc shape single crystal sample of MnSi of $\sim$15 mm in diameter and 2.15 mm in thickness with an orientation along the (100) plane was used.

Two sample setups were employed, where the propagation vector $k$ of the ultrasound wave was parallel or perpendicular to the direction of magnetic field. The propagation vector was always perpendicular to the (100) plane. The magnetic field was directed parallel or perpendicular to the (100) plane. As a result, the demagnetization factors for both setups appeared to be quite different with the smaller value in the “perpendicular“ setup, where the magnetic field was perpendicular to the propagation vector.

Details of the sample preparation were described in~\cite{8}. The LiNbO$_3$ transducers were bonded to the sample with silicon grease. The temperature was measured using a calibrated Cernox sensor with an accuracy of 0.02 K.

A sinusoidal pulse of $\sim$40 MHz was sent to the transducer that excited a sound wave inside the sample, which experienced multiple reflections. The sound travel time was determined by performing a cross-correlation between two selected reflections. The speed of sound and elastic constants are then calculated using the known thickness and density of the samples and the relationship $c_{ij}=\rho V^2$. The precision of the sound velocity determinations is no worse than one part in $\sim10^6$ whereas the absolute accuracy is about 0.5\%. The sound attenuation is estimated from an analysis of the amplitudes of two successive reflections as $20/L_0log_{10}(A_1/A_2).$

\begin{figure}[htb]
\includegraphics[width=80mm]{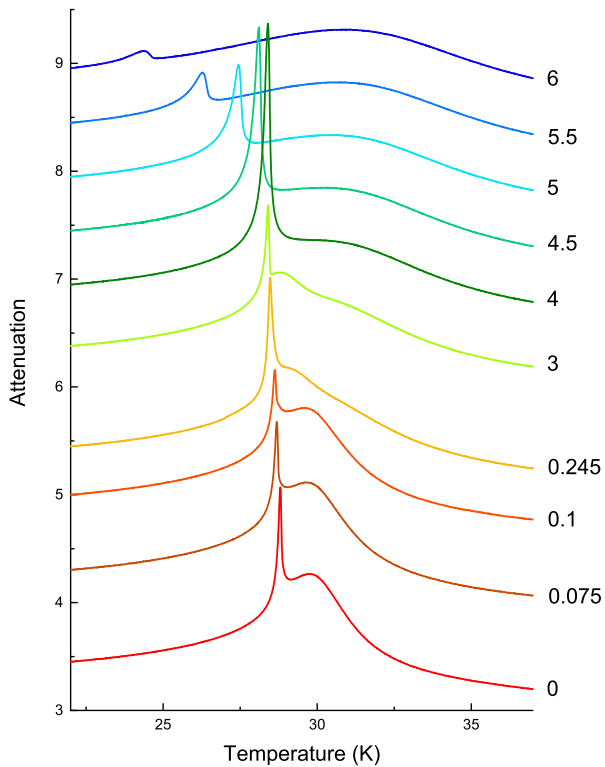}
\caption{\label{fig3} (Color online) Absorption of longitudinal sound waves at the phase transition in MnSi as functions of magnetic field and temperature, $k$, $u$, $H\parallel[001]$ (the data are shown with offsets for better viewing). The numbers in the plot stand for values of magnetic field in Tesla.}
 \end{figure}
 
\begin{figure}[htb]
\includegraphics[width=80mm]{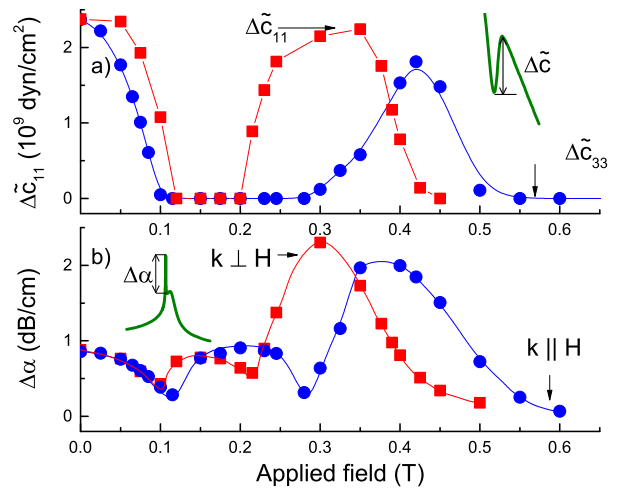}
\caption{\label{fig4} (Color online) a) Amplitudes of $\widetilde{c}_{11}$ and $\widetilde{c}_{33}$ discontinuities at the phase transition in MnSi as functions of magnetic field. Inset in the upper right corner shows a way of determining the amplitudes. b) Partial amplitudes of absorption of longitudinal sound waves at the phase transition in MnSi as functions of magnetic field. Inset in the upper left corner shows a way of determining the amplitudes. The perpendicular setup $k\perp H$ corresponds to the smaller demagnetization factor. The difference between both sets of data a) and b) in Fig.~\ref{fig4} reflect mostly the difference in demagnetization factors for two positions of our disc like sample in regards to magnetic field. That was seen already in Fig.~\ref{fig2} and would also be seen in all subsequent figures.}
 \end{figure}
 
 \begin{figure}[htb]
\includegraphics[width=80mm]{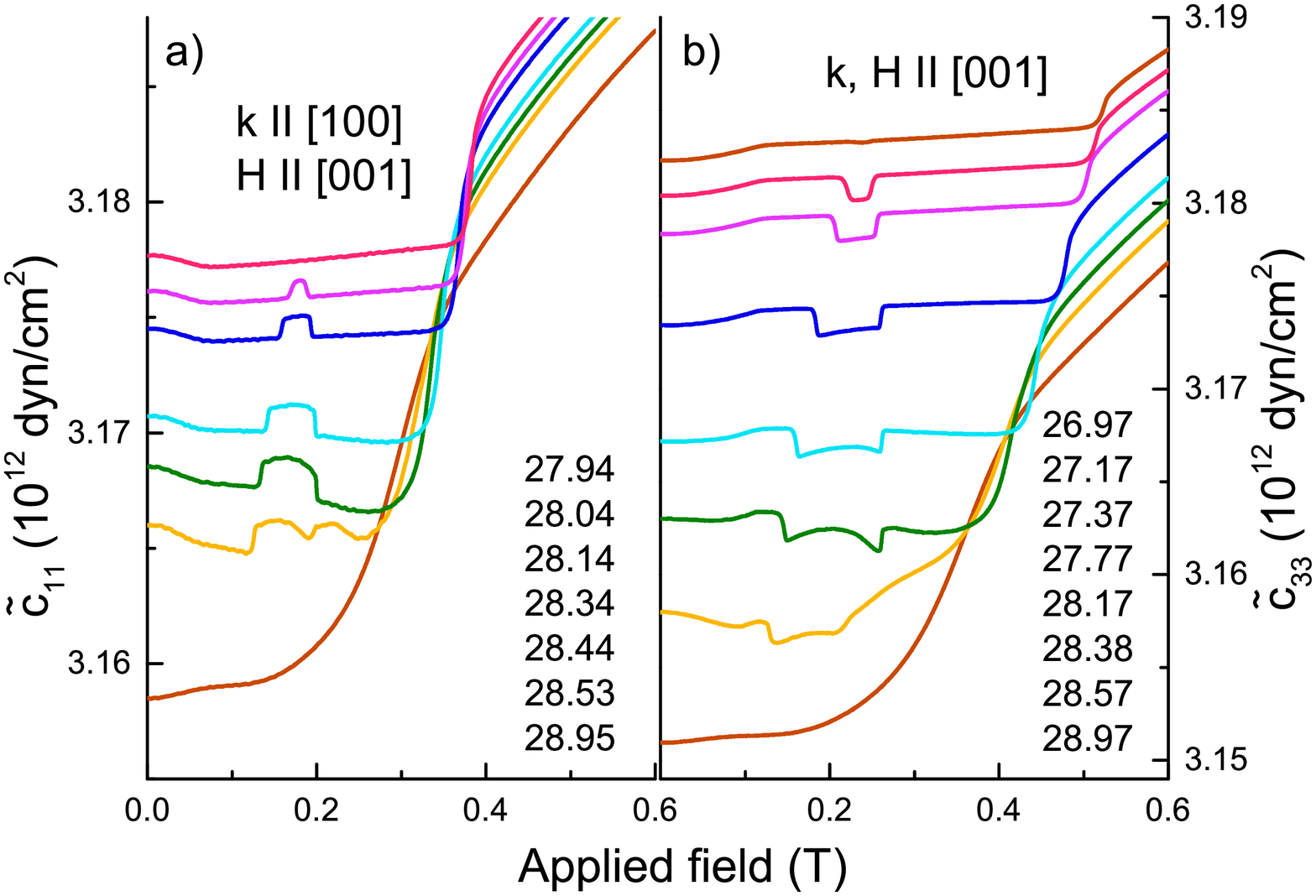}
\caption{\label{fig5} (Color online) Magnetic field dependence of the longitudinal elastic moduli at different temperatures. Temperature values for each isotherm starting from the top to the bottom are shown in the right lower corners of a) and b).}
 \end{figure}
 
 \begin{figure}[htb]
\includegraphics[width=80mm]{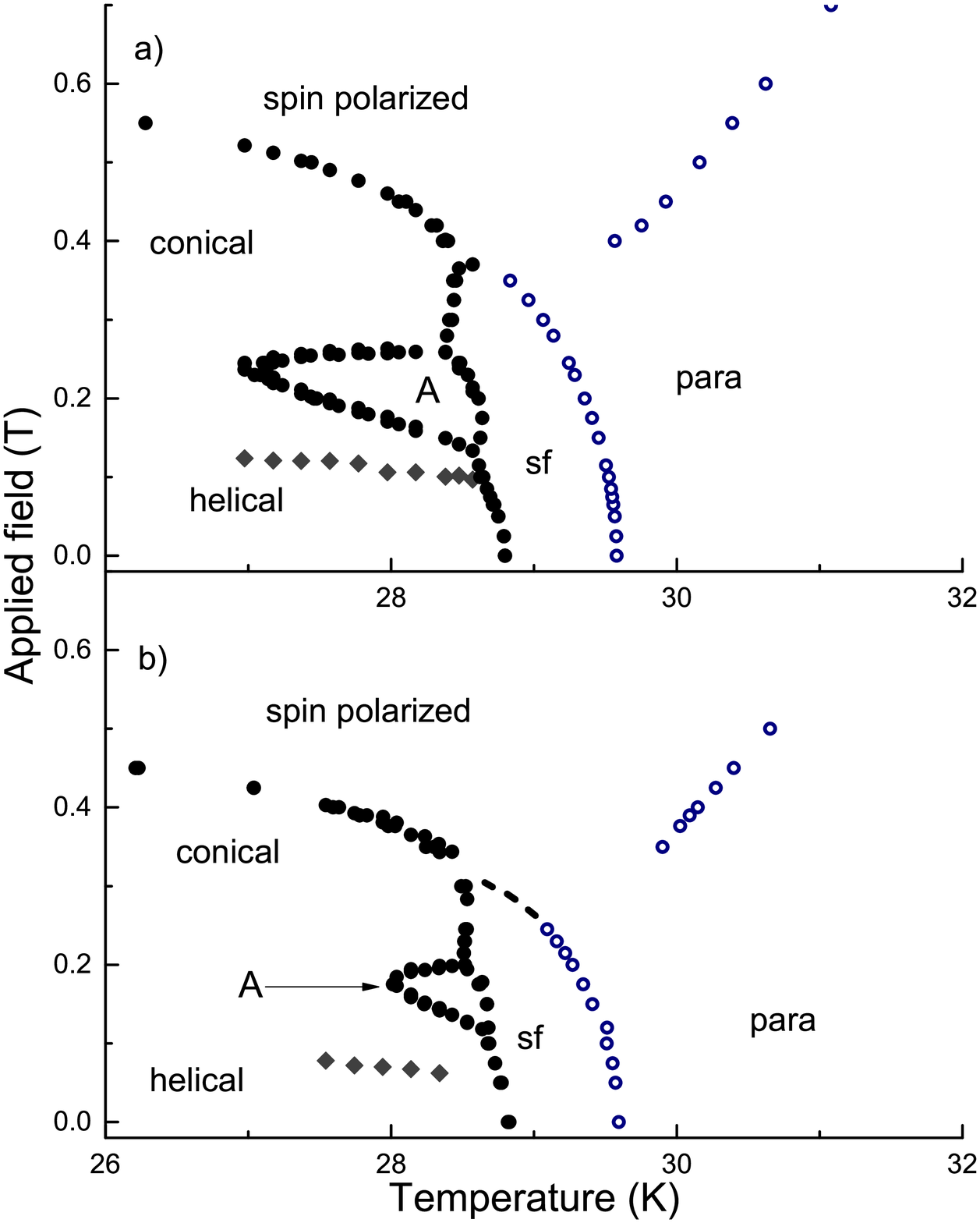}
\caption{\label{fig6} (Color online) Magnetic phase diagram of MnSi according to the present studies. Closed circles--discontinuities of $\widetilde{c}_{11}$ and $\widetilde{c}_{33}$ and maxima sound adsorption. Open circles--minima in $\widetilde{c}_{11}$ and $\widetilde{c}_{33}$, Diamond--helical=conical transition, sf--strongly fluctuating region.
}
 \end{figure}

\subsection{Longitudinal wave experiments}	
The experimental data are presented in Figs~\ref{fig2}{--}\ref{fig9}. The temperature dependence of the elastic moduli $\widetilde{c}_{11}$, $\widetilde{c}_{33}$ and corresponding data on the sound attenuation as a function of applied magnetic fields are shown in Figs.~\ref{fig2}, \ref{fig3} and \ref{fig4}. Note that application of a magnetic field to a cubic crystal makes it anisotropic. Accordingly we treat MnSi in a magnetic field directed along [001] as a tetragonal crystal (see also~\cite{16}). However, to avoid confusion we prefer to use the tilde sign to mark the corresponding elastic moduli to distinguish them from the constants of real tetragonal symmetry.

Fig.~\ref{fig5} displays the field dependence of $\widetilde{c}_{11}$ and $\widetilde{c}_{33}$ at various temperatures. Distinct discontinuities of different signs at of the border of the conical and skyrmion phases most probably suggest a first order phase transition~\cite{16}. It should also be pointed out that the elastic anisotropy, shown in Fig.~\ref{fig5}, generally agrees with the data of Ref.~\citep{16}, and corresponds to the uniaxial symmetry induced by magnetic field.

The magnetic phase diagram of MnSi obtained as a result of these measurements is demonstrated in Fig.~\ref{fig6}. Note that the values of magnetic fields corresponding to some characteristic features of the diagram obtained in the two configurations, with magnetic field either parallel or perpendicular to the direction of sound wave propagation, are different. However, this situation is obviously connected to a difference in demagnetization factors arising due to the disc shape of the sample, when the magnetic field is directed parallel or perpendicular to the disk plane~\cite{18}.

As is seen in Figs.~\ref{fig2} and \ref{fig4} the magnetic phase transition in MnSi in zero magnetic field is signified by a quasi discontinuity in $c_{11}$, which suggests a first order character of the transformation.

It is seen from Figs.~\ref{fig2} and \ref{fig4} that the quasi discontinuities $\Delta\widetilde{c}_{11}$ and $\Delta\widetilde{c}_{33}$ initially strongly decrease with magnetic field, practically disappear in the range of $\sim$0.1-0.2 ($\sim$0.1-0.3) T, then recovers at $\sim$0.2-0.45 ($\sim$0.3-0.5) T and again vanished at higher magnetic field. The numbers correspond to the measurements performed in the parallel and perpendicular setups, respectively. The method to evaluate the amplitude $\Delta\widetilde{c}_{11}$ and $\Delta\widetilde{c}_{33}$ is illustrated in the inset of Fig.~\ref{fig4}.

This kind of evaluation of $\Delta\widetilde{c}_{11}$ and $\Delta\widetilde{c}_{33}$ does not account for anomalous behavior different from quasi discontinuities that also can be seen in Fig.~\ref{fig2}. Note that the region with almost zero values of $\Delta\widetilde{c}_{11}$ and $\Delta\widetilde{c}_{33}$ in the interval $\sim$0.12-0.26 ($\sim$0.12-0.2) T closely corresponds to the extent of the skyrmion phase along the magnetic to paramagnetic transition (Fig.~\ref{fig6}).

This implies that the $\widetilde{c}_{11}$ and $\widetilde{c}_{33}$ elastic constants are almost continuous through the transition from the skyrmion to the paramagnetic phase. Another notable feature of Fig.~\ref{fig4} are the maxima of $\Delta\widetilde{c}_{ii}$ and the attenuation at $\sim$0.4(0.3) T, which correspond to the position of the announced tricritical point~\cite{15}. The data on the sound attenuation (Figs.~\ref{fig3}, \ref{fig4}b) confirm a separation of the boundary between paramagnetic and magnetically ordered phases into three distinct regions, which can be named as helical, skyrmionic, and strongly fluctuating. It needs to be emphasized that the positions of the minima in $\widetilde{c}_{11}(T)$ and $\widetilde{c}_{33}(T)$ shift with magnetic field (Fig.~\ref{fig2}) in such a way that they meet the phase transition line at $\sim$0.4 (0.3) T (Fig.~\ref{fig6}).

  \begin{figure}[htb]
\includegraphics[width=80mm]{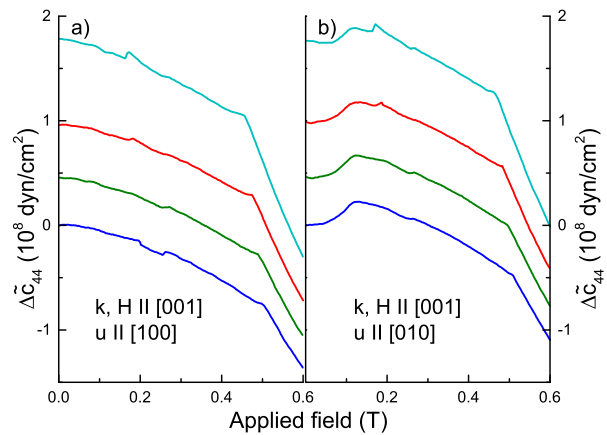}
\caption{\label{fig7} (Color online) Shear modulus $\widetilde{c}_{44}$, measured in the (001) plane with polarization $u$ along the [100] and [010] directions. Here $\Delta\widetilde{c}_{ii}={c}_{ii}-{c}_{ii}(H=0, 27.45 K)$. The temperatures of the isotherms, from top to bottom, correspond to T in K: 28.07, 27.85, 27.64, 27.45.}
 \end{figure}

 \begin{figure}[htb]
\includegraphics[width=80mm]{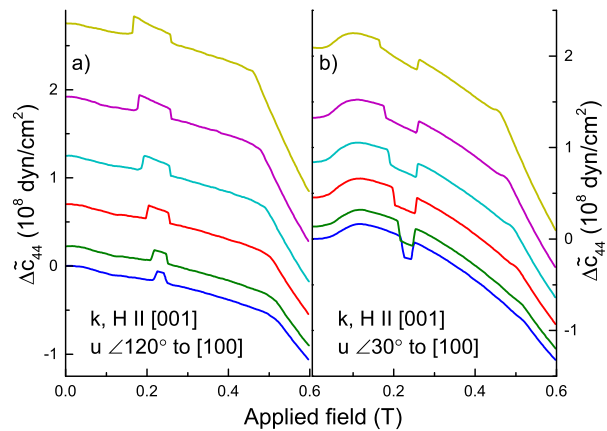}
\caption{\label{fig8} (Color online) Shear modulus $\widetilde{c}_{44}$, measured in the (001) plane with polarization $u$ at an angle of 30 and 120 degrees to the [100] direction. Here $\Delta\widetilde{c}_{ii}={c}_{ii}-{c}_{ii}(H=0, 27.14 K)$. The temperatures of the isotherms, from top to bottom, correspond to T in K: 28.06, 27.84, 27.64, 27.44, 27.24, 27.14.}
 \end{figure}

 \begin{figure}[htb]
\includegraphics[width=80mm]{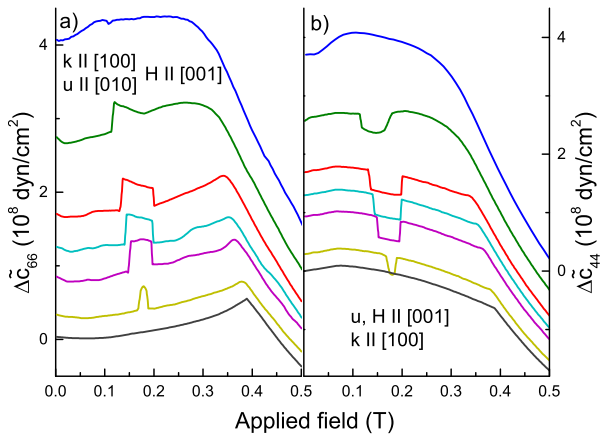}
\caption{\label{fig9} (Color online) Shear moduli $\widetilde{c}_{44}$ and $\widetilde{c}_{66}$ measured in the (100) plane. Here $\Delta\widetilde{c}_{ii}={c}_{ii}-{c}_{ii}(H=0, 27.94 K)$. The temperatures of the isotherms, from top to bottom, correspond to T in K: 28.85, 28.63, 28.43, 28.33, 28.24, 28.04, 27.94.}
 \end{figure}

\subsection{Shear wave experiments}

Experimental results on the propagation of the shear wave in MnSi are displayed in Figs.~\ref{fig7}{--}\ref{fig9}. Fig.~\ref{fig7} shows the shear elastic response of MnSi with both the propagation vector $k$ and magnetic field $H$ directed along [001], whereas polarization vector $u$ is directed along [010] and [100].

One should remember that in the case of $H=0$ this response corresponds to the $c_{44}$ elastic constant even at an arbitrary direction of polarization $u$. In the current case of $H\neq0$ a well-expressed anisotropy is quite evident.

We practically cannot see the skyrmions at the indicated directions of polarization (Fig.~\ref{fig7}). Surprisingly, however, at some special positions skyrmion features are clearly exposed (Fig.~\ref{fig8}). In contrast, in the perpendicular setup of $H\parallel[001]$ and $u\parallel[100]$ or [010] the skyrmions are clearly seen, see Fig.~\ref{fig9}.

\section{Discussion}
As was pointed out in the Introduction the heat capacity, thermal expansion coefficient, and the temperature coefficient of the electrical resistivity of MnSi at zero magnetic field are characterized by sharp peaks situated on the low temperature side of shallow maxima or minima~\cite{6,7}. The sharp peaks are considered now to be revelations of first order phase transitions, whereas the shallow maxima or minima arise as a result of intense helical fluctuations~\cite{9,19}. The peaks in the above-mentioned quantities are the result of differentiations of step like functions, which describe the behavior of enthalpy, volume, and resistivity at first order phase transitions. At the same time the sound wave propagations directly probe values of elastic moduli and immediately give values of any discontinuities in elastic properties. Therefore, the behavior of elastic moduli inferred from ultrasound studies better characterizes the nature of phase transitions than heat capacity or thermal expansions measurements. The latter can be misleading when, for instance, a first order phase transition smears out by impurities or mechanical stresses.

In the present case (see Figs.~\ref{fig2} and \ref{fig4}) the elastic modulus $\widetilde{c}_{11}$ is practically discontinuous with a positive change $\widetilde{c}_{11,p}>\widetilde{c}_{11,h}$, ( $p$ and $h$ stand for paramagnetic and helical) through the phase transition at zero magnetic field. 
The quasi discontinuities $\Delta \widetilde{c}_{11}$ decrease with magnetic field and become nearly zero across the phase transition from the skyrmion to the paramagnetic phase, in the range $\sim$0.12-0.26 ($\sim$0.12-0.2) T.

A reasonable explanation of the observed facts would be that the paramagnetic phase coexisting with the skyrmion phase is dominated by “skyrmion” fluctuations. This implies that a skyrmion liquid with a longitudinal elasticity not much different from that of the skyrmion crystal may arise in MnSi on application of a magnetic field. Actually this could be verified in neutron scattering experiments (strong chiral skyrmion like fluctuations were detected slightly above $T_c$ at $H=0$ [9,19]). Another explanation could be connected with an inefficiency of the Brazovski mechanism of appearance of a first order phase transition in case of the skyrmion--paramagnet transformation~\cite{10}.

Another puzzle can be seen in Fig.~\ref{fig4}. As one can see, the unexpected maximum of $\Delta\widetilde{c}_{11}$ and $\Delta\widetilde{c}_{33}$, accompanied by an enhancement of the sound attenuation, is located in the vicinity of 0.4 (0.3) T, where the minimum in $\widetilde{c}_{11}(T)$ and $\Delta\widetilde{c}_{33}$ approaches the phase transition line. This minimum originates from the strong “helical” fluctuations, which are believed to induce a first order phase transition in MnSi at least at zero magnetic field. Do these fluctuations help to recover first order features of the phase transition? Probably yes, because these features along with the sound attenuation decay rapidly in high magnetic fields, when the "helical" fluctuations are finally suppressed. (Note that the minima in $\widetilde{c}_{11}(T)$ at $B>0.35$ T are connected with a transition from the spin polarized paramagnetic state to a paramagnet.). In any case, a fluctuation mechanism of a first order phase transition should stop working at low temperatures, when the thermal fluctuations die out. So if one believes that the phase transition in MnSi is first order at zero magnetic field then there should exist a corresponding tricritical point, where a first order transition meets a second order one. An existence of a tricritical point on the transition line in MnSi at $\sim$0.4 T and $\sim$28 K was claimed in Ref.~\cite{15}. Our ultrasound study cannot support this conclusion. Indeed, an elastic modulus $\widetilde{c}_{11}$ is expected to diverge at a tricritical point, which is not observed in the present investigation (see Fig.~\ref{fig2})~\cite{20}.

Taking a broader view of the situation one may come to the conclusion that the magnetic phase transition in MnSi is weak first order probably in the whole range of magnetic fields studied. Taking into account a first order nature of the phase line between conical and skyrmion phases, one may conclude that the crossing points 1,2 (Fig.~\ref{fig1}c) are usual triple points, which are not accompanied by an anomalous behavior of elastic properties. The quasi discontinuities of $\widetilde{c}_{11}$ and $\widetilde{c}_{33}$ vary from small to tiny under an influence of magnetic fields. Their recovery and the enhanced sound absorption occur at the intersection of the phase transition line and the line of minima in $\widetilde{c}_{11}$ and $\widetilde{c}_{33}$. The latter is a region of strong spin fluctuations, reflected in maxima of heat capacity, etc. Despite the fact that this region does not carry features of a real phase transition, the existing powerful fluctuations at the minima of $\widetilde{c}_{11}$ and $\widetilde{c}_{33}$ make the mentioned intersection point similar to a critical end point, where a second order phase transition meets a first order one. The best example of this situation is the crossing of the $\lambda$--transition line in liquid helium and its melting curve, which is accompanied by anomalies in the thermodynamic properties of the melting transition.

One more enigma of the present study is the discovery of anisotropic behavior of the shear modulus (see Figs.~\ref{fig7}, ~\ref{fig8}). The anisotropy was spotted at different positions of the polarization vector $u$ in the (001) plane in the parallel setup. An evidence of the helical--conical transition at $\sim$0.1 T can be seen only at some selected positions of vector $u$.

Another surprising detail is that the skyrmion phase does not reveal itself at the regular directions of the polarization vector, along [100] or [010]. Instead, it can be clearly observed at angles of $\sim$30 or 120 degrees for these directions, which certainly suggests a connection with the triangular packing. A skyrmion crystal is believed to belong to this class of lattice. But this association can not be verified in the case of true hexagonal symmetry. However, the observed anisotropy is, as a matter of fact, small. The difference of the shear moduli of the skyrmion phase measured in two directions of polarization vector is only $\sim 2\times10^{-3}$ \% (see Fig.~\ref{fig8}). Hence, we may conclude that the hexagonal symmetry of the skyrmion crystal is slightly distorted by the magnetic field inhomogeneity or structural defects, which can easily be missed in neutron studies. On the other hand, the measurements of the shear moduli in the perpendicular setup do not show any noticeable anomalies.

Getting to a discussion of the magnetic phase diagram of MnSi, we should note that the spotted scale difference in figures entitled as a and b is explained by a difference in the demagnetization factor. Indeed, in our sample the demagnetization factors differ by a factor of five for the parallel and perpendicular setups. At the same time, it needs to be emphasized that the skyrmion domain in case of the "perpendicular" setup with the smaller demagnetization factor has a reduced temperature range, which suggests that the magnetic field inhomogeneity plays an important role in the skyrmion occurrence.

\section{Conclusion}
The sound speed and attenuation were measured in a MnSi single crystal in the temperature range of 2 -- 40 K and magnetic fields to 7 Tesla.

The magnetic phase transition in MnSi in zero magnetic field is signified by a quasi discontinuity in the $c_{11}$ elastic constant, which suggests a first order character of the transformation. The quasi discontinuities of $\widetilde{c}_{11}$ and $\widetilde{c}_{33}$ initially strongly decrease with magnetic field, practically disappear in the range of $\sim 0.1-02/0.3$ T, then recover at $\sim 0.2/0.3-0.45/0.5$ T and again vanish at higher magnetic field. Note that the region with almost zero values of $\widetilde{c}_{11}$ and $\widetilde{c}_{33}$ in the  interval $\sim 0.1-0.2/0.3$ T closely corresponds to the extent of the skyrmion phase along the magnetic to paramagnetic transition. This implies that the $\widetilde{c}_{11}$ and $\widetilde{c}_{33}$ elastic constants are almost continuous through the transition from the skyrmion to the paramagnetic phase. This result probably questions the applicability of the Brazovsky theory for the skyrmion-paramagnet phase transition~\cite{10}. The recovery of the discontinuities of $\widetilde{c}_{11}$ and $\widetilde{c}_{33}$ and the enhanced sound absorption occur at the intersection of the phase transition line and the line of minima in $\widetilde{c}_{11}$ and $\widetilde{c}_{33}$. The latter is a region of strong spin fluctuations, reflected in maxima of the heat capacity. Despite the fact that this region does not carry features of a real phase transition, the existing powerful fluctuations at the minima of $\widetilde{c}_{11}$ and $\widetilde{c}_{33}$ make the mentioned intersection point similar to a critical end point, where a second order phase transition meets a first order one.

The spotted magnetic scale difference in the figures entitled a and b results from differences in the demagnetization factors. The skyrmion domain in case of the "perpendicular" setup with the smaller demagnetization factor has a reduced temperature range, which suggests that the magnetic field inhomogeneity plays an important role in the skyrmion occurrence and, hence, opens a way of skyrmion manipulation (see also~\cite{18,21}.

The elastic uniaxial symmetry of MnSi (Fig.~\ref{fig5}), induced by a magnetic field is broken by the magnetic field inhomogeneity or structural defects that also distort the hexagonal symmetry of skyrmion crystal, as evidenced by the anisotropy of the shear modulus in the (001) plane (Fig.~\ref{fig7},~\ref{fig8}).
\section{Acknowledgements}
This work was supported by the Russian Foundation for Basic Research (grant 15-02-02040), Russian Science Foundation (grant 14-22-00093) and Program of the Physics Department of RAS on Strongly Correlated Electron Systems and Program of the Presidium of RAS on Strongly Compressed Matter.  The remarks and comments of Andreas Hermann are greatly appreciated.


\begin{thebibliography}{99}
\bibitem{1}	H. J. Williams, J. H. Wernick, R. C. Sherwood, G. K. Wertheim, J. Appl. Phys. \textbf{37}, 1256 (1966).
\bibitem{2} Ishikawa, K. Tajima, D. Bloch, M. Roth, Solid State Commun. \textbf{19}, 525 (1976)
\bibitem{3} I. E. Dzyaloshinski, Zh. Eksp. Teor. Fiz. \textbf{72}, 1930 (1977)
\bibitem{4}	Per Bak, M. H{\o}gh Jensen, J. Phys. C \textbf{13}, L881 (1980)
\bibitem{5}	C. Pfleiderer, G. J. McMullan, S. R. Julian, G. G. Lonzarich, Phys. Rev. B \textbf{55}, 8330 (1997)
\bibitem{6}	S. M. Stishov, A. E. Petrova, S. Khasanov, G. Kh. Panova, A. A. Shikov, J. C. Lashley, D. Wu, T. A. Lograsso, Phys. Rev. B \textbf{76}, 052405 (2007)
\bibitem{7}	S. M. Stishov, A. E. Petrova, S. Khasanov, G. Kh. Panova, A. A. Shikov, J. C. Lashley, D. Wu, T. A. Lograsso, J. Phys.: Condens. Matter \textbf{20}, 235222 (2008)
\bibitem{8}	A. E. Petrova, S. M. Stishov, J. Phys.: Condens. Matter \textbf{21}, 196001 (2009)
\bibitem{9} C. Pappas, E. Leli\`{e}vre-Berna, P. Falus, P.M.Bentley, E. Moskvin, S. Grigoriev, P. Fouquet, B. Farago, Phys. Rev. Lett. \textbf{102}, 197202 (2009)
\bibitem{10} S. A. Brazovskii, Sov. Phys. JETP \textbf{41}, 85 (1975)
\bibitem{11} M. Janoschek, M. Garst, A. Bauer, P. Krautscheid, R. Georgii, P. B\"{o}ni, C. Pfleiderer, Phys. Rev. B 87, 134407 (2013)
\bibitem{12} Y. Ishikawa, T. Komatsubara, D. Bloch, Physica B \textbf{86-88}, 401 (1977)
\bibitem{13} Y. Ishikawa, M. Arai, J. Phys. Soc. Jpn. \textbf{53}, 2726 (1984)
\bibitem{14} S. M\"{u}hlbauer, B. Binz, F. Jonietz, C. Pfleiderer, A. Rosch, A. Neubauer, R. Georgii, P. B\"{o}ni, Science 323, 915 (2009)
\bibitem{15} Bauer, M. Garst, C. Pfleiderer, Phys. Rev. Lett. \textbf{110}, 177207 (2013)
\bibitem{16} Y. Nii, A. Kikkawa,Y. Taguchi, Y. Tokura, Y. Iwasa, Phys. Rev. Lett. \textbf{113}, 267203 (2014)
\bibitem{17} A.E. Petrova, S. M. Stishov, Instruments and Experimental Techniques \textbf{52}, 4, 609 (2009)
\bibitem{18} A. Bauer, C. Pfleiderer, Phys. Rev B \textbf{85}, 214418 (2012)
\bibitem{19} C. Pappas, E. Leli\`{e}vre-Berna, P. Bentley, P. Falus, P. Fouquet, B. Farago, Phys. Rev. B \textbf{83}, 224405 (2011)
\bibitem{20} L.D. Landau, E.M. Lifshitz, Statistical Physics, Part 1, 3rd ed., Pergamon, Oxford, 1980
\bibitem{21} X. Z. Yu, N. Kanazawa, Y. Onose, K. Kimoto, W. Z. Zhang, S. Ishiwata, Y. Matsui, Y. Tokura, Nature Materials \textbf{10}, 106 (2011)
\end{thebibliography}
\end{document}